\documentclass[9pt, conference]{IEEEtran}

\IEEEoverridecommandlockouts
\usepackage{todonotes}
\usepackage{slashbox}
\usepackage{cite}
\usepackage{amsmath,amssymb,amsfonts}
\usepackage{algorithmic}
\usepackage{graphicx}
\usepackage{textcomp}
\usepackage{xcolor}
\def\BibTeX{{\rm B\kern-.05em{\sc i\kern-.025em b}\kern-.08em
    T\kern-.1667em\lower.7ex\hbox{E}\kern-.125emX}}
\begin{document}

\title{VoxVietnam: a Large-Scale Multi-Genre Dataset for Vietnamese Speaker Recognition
% {\footnotesize \textsuperscript{*}Note: Sub-titles are not captured for https://ieeexplore.ieee.org  and
}

\makeatletter
\newcommand{\linebreakand}{%
  \end{@IEEEauthorhalign}
  \hfill\mbox{}\par
  \mbox{}\hfill\begin{@IEEEauthorhalign}
}
\makeatother

\author{\IEEEauthorblockN{Hoang Long Vu}
\IEEEauthorblockA{\textit{Hanoi University of}\\
\textit{Science and Technology}\\
Hanoi, Vietnam \\
longvu200502@gmail.com
}
\and
\IEEEauthorblockN{Phuong Tuan Dat}
\IEEEauthorblockA{\textit{Hanoi University of}\\ \textit{Science and Technology} \\
Hanoi, Vietnam \\
phuongtuandat2915@gmail.com}
\and
\IEEEauthorblockN{Pham Thao Nhi}
\IEEEauthorblockA{\textit{Hanoi University of}\\ \textit{Science and Technology }\\
Hanoi, Vietnam \\
nhi.phamt2002@gmail.com}
\linebreakand
\IEEEauthorblockN{Nguyen Song Hao}
\IEEEauthorblockA{\textit{Hanoi University of Science and Technology} \\
Hanoi, Vietnam \\
wednesdayhao@gmail.com}
\and
\IEEEauthorblockN{Nguyen Thi Thu Trang$^{*}$\thanks{\textsuperscript{*}Corresponding author}}
\IEEEauthorblockA{\textit{Hanoi University of Science and Technology} \\
Hanoi, Vietnam \\
trangntt@soict.hust.edu.vn}
}

\maketitle

\begin{abstract}
Recent research in speaker recognition aims to address vulnerabilities due to variations between enrolment and test utterances, particularly in the multi-genre phenomenon where the utterances are in different speech genres. Previous resources for Vietnamese speaker recognition are either limited in size or do not focus on genre diversity, leaving studies in multi-genre effects unexplored. This paper introduces VoxVietnam, the first multi-genre dataset for Vietnamese speaker recognition with over 187,000 utterances from 1,406 speakers and an automated pipeline to construct a dataset on a large scale from public sources. Our experiments show the challenges posed by the multi-genre phenomenon to models trained on a single-genre dataset, and demonstrate a significant increase in performance upon incorporating the VoxVietnam into the training process. Our experiments are conducted to study the challenges of the multi-genre phenomenon in speaker recognition and the performance gain when the proposed dataset is used for multi-genre training.

\end{abstract}


\section{Introduction}
% - What is speaker recognition?

% - Challenges in speaker recognition?

% - Multi-genre challenges?

% - Current progress of Vietnamese SR? (lack of genre diversity, limitations in collection pipeline)

% - Outline of the paper
Speaker recognition is a biometric technology that identifies or verifies a speaker based on their speech, with various applications in security, authentication, and forensic investigations. The development of recognition models based on Deep Neural Networks (DNNs) \cite{desplanques2020ecapa, nagrani2020voxceleb, chen2022wavlm} architectures has led to a surge in performance, allowing various commercial applications for speaker recognition systems.

Current research aims to address vulnerabilities due to intrinsic (speaking style, physiological status) and extrinsic (recording device, environment noise) variations between enrolment and test utterances, particularly with multi-genre scenarios being one of the most challenging owing to the involvement of complex variations. The authors in \cite{li2022cn} proposed the first multi-genre dataset in speaker recognition and observed an increase of up to five times in Equal Error Rate (EER) when a model trained on a single-genre dataset is tested on the multi-genre dataset. Various works \cite{zhou2024investigation, zhang2023meta, kang2022investigation} investigated the multi-genre challenges in speaker recognition. New models and training paradigms like meta-learning for speaker verification \cite{zhang2023meta} or distribution alignment \cite{zhou2024investigation} are also proposed to improve the performance in these cases. 

Nevertheless, the success in the multi-genre scenarios of DNN-based models is highly reliant on the quality and the abundance of training data. A few multi-genre datasets \cite{fan2020cn, li2022cn} have been published to facilitate the development of such models. In the speaker recognition task of low-resource languages like Vietnamese, obtaining such datasets with rich genre diversity is challenging. Currently available datasets in Vietnamese speaker recognition are either inadequate in terms of size and scale (ZaloAI, VLSP 2021 \cite{dat2022vlsp}), or lack the diversity of genre in the utterances of speakers \cite{thanh2023vietnam, thanh2024robust, hoang24b_interspeech}, leaving the performance in practical multi-genre scenarios of state-of-the-art models unexplored. Our previous work of building the largest dataset for Vietnamese, Vietnam-Celeb \cite{thanh2023vietnam} requires a manual selection of speakers in advance of the curation process. Therefore, this pipeline limits the ability to scale to a larger number of speakers.

In this paper, we propose a novel construction pipeline utilising deep clustering and multi-modal cleansing techniques and apply it to build VoxVietnam. The pipeline is not limited to the initial list of speakers, therefore can be scaled and adapted to any number of speakers and utterances, in any language. The dataset comprises 261 hours of 187,980 utterances from 1,406 speakers, covering the three most common genres in casual conversations. Using this dataset, we conduct extensive experiments to: (i) validate the quality of the dataset, (ii) investigate the challenges posed by the real-world multi-genre scenario, and (iii) examine the performance gain from using VoxVietnam as the training source. We believe that VoxVietnam is a valuable resource to facilitate further research on the impacts of multi-genre challenges in speaker recognition.

The remaining sections of this paper are organised as follows: Section 2 describes the proposed data construction pipeline, Section 3 introduces the resultant dataset from the pipeline, and Section 4 discusses the related works. Finally, the experimental setup and results are detailed in Section 5, and we draw our conclusions in Section 6.

\section{Proposed Data Construction Pipeline}
This section will describe the proposed construction pipeline and how each step can be utilised to build the dataset. 

\subsection{Proposed Pipeline}
\begin{figure}[h]
    \centering
    \includegraphics[width=\linewidth]{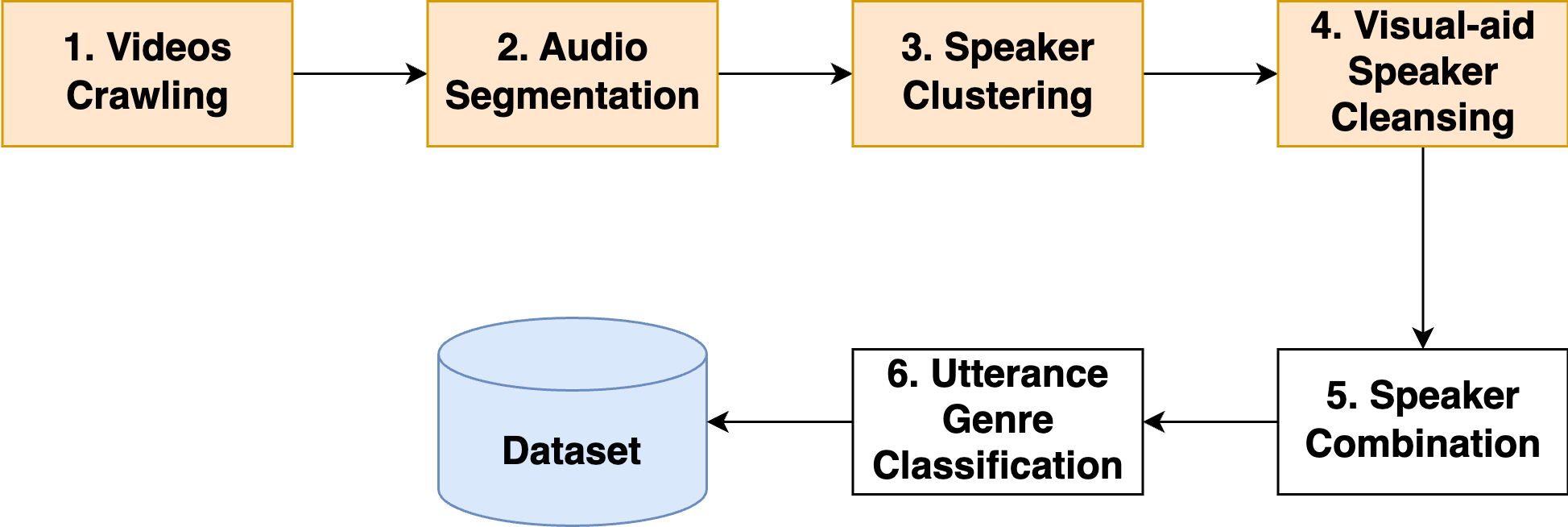}
    \caption{An overview of the proposed data construction pipeline.}
    \label{fig:overview-pipeline}
\end{figure}
The overall data construction pipeline is illustrated in Figure \ref{fig:overview-pipeline}, consisting of six main steps. The steps denoted in orange blocks are done for each playlist. The expected outcome of the pipeline is a dataset comprising utterances grouped by speaker and genre.

Firstly, we manually list a set of keywords that describe various types of activities or genres, then automatically search for playlists with titles matched with each of the keywords. After that, we apply the proposed pipeline to the collected playlist URLs.

\textbf{1. Videos Crawling.}   In this step, all videos in a playlist are crawled. We only collect videos of at least 480p in quality and uploaded after 2017 to ensure sufficient quality for processing steps.

\textbf{2. Audio Segmentation.} We extract the audio from all videos in the playlist, then remove the silence regions by a simple Voice Activity Detection (VAD) model. Since the audio may contain speeches from more than one speaking person, we further utilise a pre-trained Speaker Diarisation model to segment into speaker-homogeneous segments.

\textbf{3. Speaker Clustering.} With the help of a pre-trained speaker encoder in Vietnamese, we extract the speaker embeddings from segmented audios to prepare for the multiple-staged speaker clustering. The outcome of this step is utterances grouped by speakers based on the clustering the speaker embeddings. More details on this step are explained in Section \ref{sec:clustering}.

\textbf{4. Visual-aid Speaker Cleansing.} Since the speaker clustering only observes the information on the audio modality, we apply a visual-aid cleansing step to eliminate utterances with noisy speaker labels. More details in this step are covered in Section \ref{sec:cleansing}.

\textbf{5. Speaker Combination.} Speakers can appear in various collected playlists. Therefore, after applying Steps 1-4 for every playlist we merge speakers with high cosine similarity on both audio and visual modality. 

\textbf{6. Utterance Genre Classification.} Since the genre information in the video metadata in the crawling step may not reflect exclusively for each utterance in the video, we classify the genre of each utterance using a DNN-based model. Further details are provided in Section \ref{sec:genre-classify}

\subsection{Speaker Clustering}
\label{sec:clustering}
We borrow the idea from VoxTube \cite{yakovlev2023voxtube} and VoxCeleb2 \cite{chung2018voxceleb2}, with the hypothesis that most of the YouTube channels have one predominantly person speaking. Therefore, a comprehensive pipeline of clustering algorithms can help to find the utterances of the predominant speaker, which was applied successfully in \cite{yakovlev2023voxtube}. However, our work relaxes the hypothesis to: Each YouTube playlist has \textit{at least one} predominant speaker, so as to fully exploit the collected videos. We also propose several modifications to the original pipeline in \cite{yakovlev2023voxtube} to group utterances by speakers in the case of multi-speaker playlists.

Specifically, speaker diarisation was utilised to first segment the original audio into speaker-homogeneous utterances. The clustering steps in Speaker Clustering in a playlist are similar to the scheme of the filtration pipeline for one channel in \cite{yakovlev2023voxtube}. We use DBSCAN \cite{deng2020dbscan} clustering algorithm rather than the Hierarchical Agglomerative Clustering (HAC) to mitigate the sensitivity to noisy samples. We also keep all of the clusters apart from noisy clusters rather than keeping only the largest one as implemented in the original work.

\subsection{Visual-aid Speaker Cleansing}
\label{sec:cleansing}
\begin{figure}[]
    \centering
    \includegraphics[width=\linewidth]{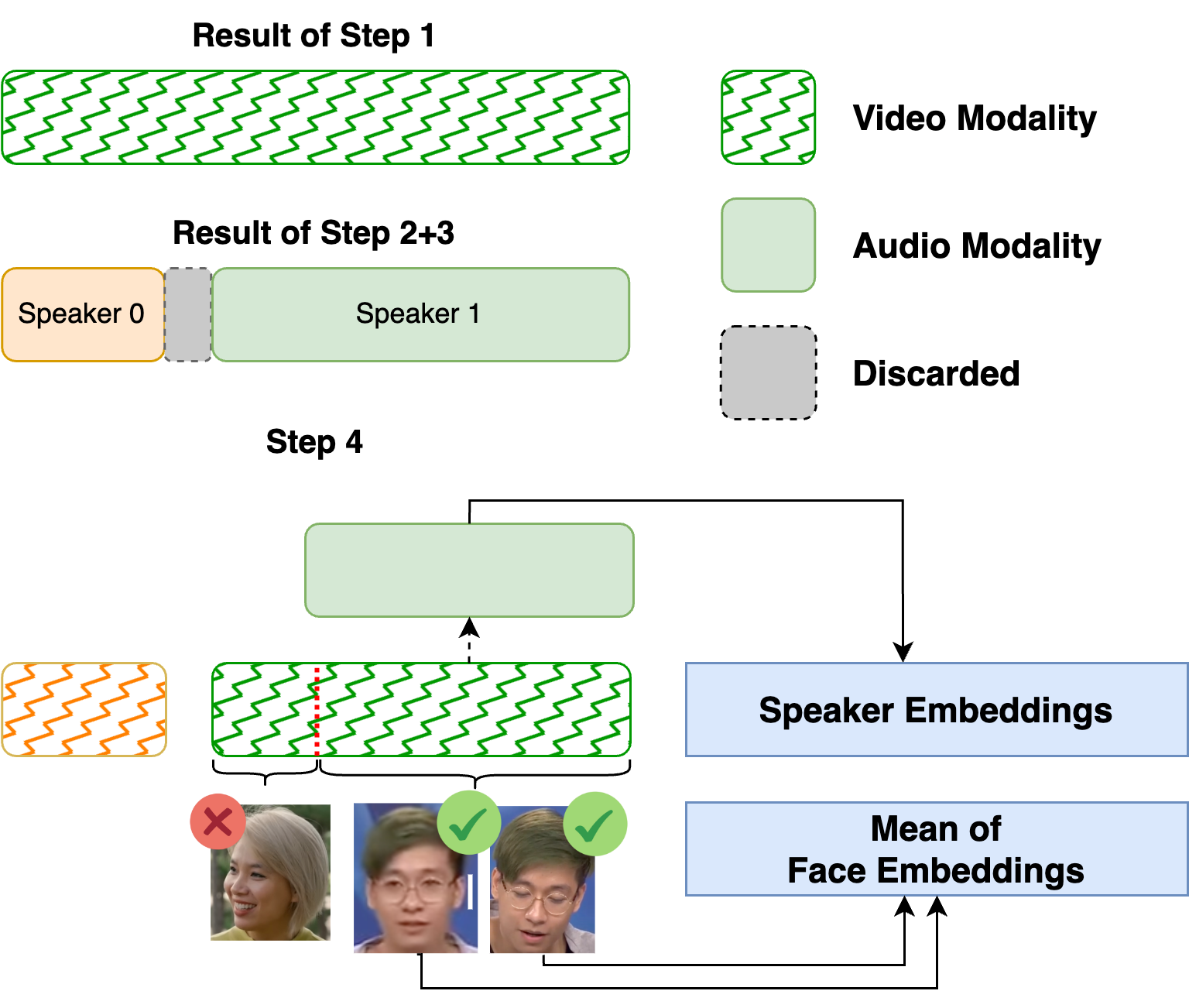}
    \caption{The speaker cleansing pipeline for each utterance.}
    \label{fig:cleansing}
\end{figure}
The authors in \cite{tao20b_interspeech} argued that visual modality can help to detect and remove samples that are falsely assigned to a speaker in automatically curated datasets. Therefore, we propose an automatic cleansing procedure, combining visual and audio modalities to decide if an utterance of a speaker is noisy or not. The proposed pipeline is illustrated in Figure \ref{fig:cleansing}.

For each utterance belonging to a cluster (speaker), we extract the corresponding video segment from the original video collected in Step 1 based on the timestamps. After that, we employed S\textsuperscript{3}FD \cite{zhang2017s3fd} for face tracking and TalkNet \cite{beliaev21_interspeech} for active speaker verification for each frame after 100 frames. The detected frames corresponding to the speaking person are extracted and cropped to their face only. After that, the face embeddings of detected frames are aggregated by mean. 

Next, the distance from the considered utterance to all other ones in the same cluster is computed in terms of speaker embedding distance and face embedding distance, then normalised by the cluster size. Finally, the harmonic mean of these distances is computed and compared to a threshold to decide whether or not the considered utterance really belongs to the speaker or not. We set a conservative threshold of 0.75 in this work to ensure the highest quality of the automated pipeline. 
% The proposed method is mathematically described as follows:
% \begin{align*} 
%     x_i = \frac{1}{M_k}\sum_{j=1}^N I_{i=j}cos(s_i, s_j)\\
%     y_i=\frac{1}{M_k}\sum_{j=1}^N I _{i=j}cos(f_i, f_j)\\ \\
%     z_i = \frac{2x_iy_i}{x_i+y_i}\\
% \end{align*}

% Here $M_k$ is the number of samples in the $k^{th}$ cluster, $I$ is an indicator function evaluating 1 when $i = j$.

\subsection{Utterance Genre Classification}
\label{sec:genre-classify}
After clustering and cleansing, we use the Audio Spectrogram Transformer (AST) model \cite{gong21b_interspeech} to classify the genre of the utterance. This is the state-of-the-art model in audio classification task using Google SpeechCommand V2 dataset \cite{Warden2018SpeechCA}. The AST model is based on the architecture of Vision-Transformer model \cite{dosovitskiy2021imageworth16x16words}, pre-trained on ImageNet \cite{5206848} and AudioSet \cite{7952261} dataset. We need to fine-tune the model on Vietnamese utterances to classify the genre.

Our dataset for fine-tuning the AST model on Vietnamese comprises 281.35 hours of short utterances, whose construction progress is described as follows. First, we crawl videos from YouTube, which are manually found based on titles containing keywords related to our pre-defined genre labels: spontaneous, reading, and singing. The videos are then processed to extract audio files, and a VAD model is applied to segment the long audio files into smaller segments and discard non-speech signals. We divide the videos into training, validation, and evaluation subsets. The videos each subset are ensured not to be crawled from the same channel to avoid introducing bias in the evaluation. The number of utterances in training/validation/evaluation subsets is 233,754 / 40,882 / 23,030 respectively. The genre of utterances in the test subset is manually annotated by our team. The F1 score on the spontaneous, reading, and singing classes during evaluation with the AST model is 0.94, 0.89, and 0.77 respectively.

% \begin{table}[h]
% \centering
% \caption{Performance of genre classification model}
% % \resizebox{\columnwidth}{!}{%
% \begin{tabular}{|lr|}
% \hline
% \textbf{Genre} & \textbf{F1 score} \\ \hline \hline
% Spontaneous &  0.94  \\
% Reading & 0.89 \\
% Singing & 0.77 \\
% \hline
% \end{tabular}
% \label{tab:genre-class}
% \end{table}

\section{The VoxVietnam Dataset}
After applying the proposed pipeline, we obtain the VoxVietnam dataset with 187,980 utterances from 1,406 speakers, totalling 261 hours. All utterances are resampled to 16,000Hz. Our dataset covers a wide range of challenging scenarios, from interviews, and gameshow to podcasts and entertainment videos. VoxVietnam is the largest dataset for Vietnamese speaker recognition, covering the three most common genres in real-world scenarios: reading speech, spontaneous speech, and singing. This section will describe several statistics of VoxVietnam.
\subsection{Utterance and Genre Distribution}
% The total number of crawled videos by each keyword are demonstrated in Table \ref{tab:crawled-videos}. Our collection sources are diverse, covering most of the challenging scenarios in the wild to support the development of robust recognition models.
Table \ref{tab:duration-dist} shows the utterance length distribution of the
dataset. Short utterances of under 5 seconds make up a high amount of our data, which represents the real-world speaker recognition task, where the audio inputs are mostly short. Moreover, the distribution also reflects the spontaneous nature of speech in practice.
\begin{table}[h]
\centering
\caption{Utterance duration distribution VoxVietnam}
% \resizebox{\columnwidth}{!}{%
\begin{tabular}{|rrr|}
\hline
\textbf{Duration (s)} & \textbf{\# of Utterances} & \textbf{Proportion (\%)} \\\hline\hline
\textless{}2          & 44,288                    & 23.56                     \\
\textbf{2-5}          & \textbf{95,932}           & \textbf{51.03}            \\
5-10                  & 23,426                    & 12.46                     \\
10-20                 & 19,125                    & 10.17                    \\
\textgreater{}20      & 5,209                    & 2.78                     
\\\hline
\end{tabular}%
% }
\label{tab:duration-dist}
\end{table}

Table \ref{tab:genre-dist} illustrates the proportion that each genre accounts for. It can be seen that the genre distribution matches the nature of resources from public media: mostly spontaneous, accounting for over 80\%. The reading genre also makes up a significant proportion. On the other hand, singing only accounts for about 3\%, totalling over 5,700 utterances. The severe imbalance may be due to three factors: (i) the number of crawled videos by the keyword related to singing is insufficient in the first place, (ii) the total duration of singing videos is generally not long, since music videos are mainly composed of one or a few songs, and (iii) the speaker embedding extractor is not trained on singing utterances, therefore the clustering algorithms could not achieve good results in distinguishing these utterances from outliers.

\begin{table}[]
\centering
\caption{The genre distribution of VoxVietnam}
\label{tab:genre-dist}
\begin{tabular}{|lrrr|}
\hline
\textbf{Genre} & \textbf{\# of Spks} & \textbf{\# of Utterances} & \textbf{\# of Hours} \\ \hline\hline
Spontaneous    & 1,261                   & 150,513                   & 207.82               \\
Reading        & 717                     & 31,702                    & 41.80                \\
Singing        & 545                     & 5,765                     & 11.91                \\ \hline\hline
\textbf{Total} & \textbf{2,523}          & \textbf{187,980}          & \textbf{261.53}      \\ \hline
\end{tabular}
\end{table}
\subsection{Comparison with other datasets}
Table \ref{tab:compare-other} shows the comparison between the proposed dataset with other datasets in Vietnamese (below), and other public multi-genre datasets in the world (above), datasets with names in \textit{italics} are multi-genre. It can be seen that VoxVietnam is the largest and the only multi-genre dataset for Vietnamese speaker recognition, which will be a valuable resource for developing models in the complex nature of real-world speeches.

\begin{table}[]
\centering
\caption{Comparison with other datasets}
\label{tab:compare-other}
\begin{tabular}{|lrrr|}
\hline
\textbf{Name}       & \textbf{\# of Spks} & \textbf{\# of Utterances} & \textbf{\# of Hours} \\ \hline\hline
\textit{CN-Celeb1}           & 1,000                   & 130,109                   & 274                  \\
\textit{CN-Celeb2}           & 2,000                   & 529,485                   & 1,090                \\ \hline\hline
VLSP 2021           & 1,305                   & 31,600                    & 41                   \\
Vietnam-Celeb       & 1,000                   & 87,140                    & 187                  \\
\textbf{\textit{VoxVietnam}} & \textbf{1,406}            & \textbf{187,980}          & \textbf{261}         \\ \hline
\end{tabular}
\end{table}

\section{Related Works}
The largest and most commonly used dataset for benchmarking in Vietnamese speaker recognition task is our previous work in Vietnam-Celeb \cite{thanh2023vietnam}. However, the number of speakers and utterances are still limited compared to other datasets in high-resource languages and does not takes the genre diversity into account, leading to suboptimal performance in complex multi-genre scenarios. Moreover, the Vietnam-Celeb collection pipeline requires a manual listing of speakers in advance, which can be time-consuming and hard to extend to a larger scale.

VoxTube \cite{yakovlev2023voxtube} was the first work to propose a multi-staged clustering schema to automatically group the utterances of the same speaker together. However, we relax the hypothesis proposed in VoxTube and further introduce several steps to ensure the highest accuracy of the utterances. 

To further enhance the quality of the dataset, we take the visual information into consideration to clean noisy samples after clustering. Specifically, an utterance of a speaker are only considered valid if it has high speaker and face similarity with all other utterances. This proposal was motivated by improved recognition results and a reduced number of noisy samples after introducing visual information to eliminate the effect of noisy labels in speaker representation learning \cite{tao20b_interspeech}.

\section{Experiments}
% \subsection{Experiment Questions}
Our experiments focus on investigating the challenges posed by the multi-genre scenarios, and the effectiveness of our proposed dataset for the training purpose. Besides, we also conduct experiments to compare the recognition performance in different genres.
\subsection{Experimental Setup}
We choose the ECAPA-TDNN architecture \cite{desplanques2020ecapa} for the experiments. The model is optimised with the AAM-Softmax loss \cite{deng2019arcface}. Every experiment is trained for 100 epochs on one NVIDIA Tesla V100 32GB GPU with the hyperparameter and optimiser chosen similar to the original paper \cite{desplanques2020ecapa}. For data augmentation, we use the MUSAN dataset \cite{snyder2015musan} and Room Impulse Response simulation \cite{ko2017study}. The evaluation metric used in the experiments is Equal Error Rate (EER). 

We first divide the original dataset into a training set and two test sets, denoted VoxVietnam-T, VoxVietnam-E, and Vox-Vietnam-H respectively. The speakers between the training and two test sets are disjoint. We also introduce VoxVietnam-T-\textit{small} with roughly the same number of speakers and utterances as in the Vietnam-Celeb training set to show the unbiased training effectiveness of our proposed data. Besides, to investigate the performance gain of the proposed speaker cleansing step in the pipeline, we also create VoxVietnam-T-\textit{noisy} with utterances before being cleansed. 

The authors in \cite{li2022cn} argued that the speakers with speech utterances of only one genre, or utterances collected from multiple sessions in diverse conditions are still valuable to multi-genre performance. Note that in this case, we treat each video as a single session. To further evaluate the contributions of genre-isolated speakers (speakers with different genres considered as different speakers) and session-isolated speakers (speakers in different sessions considered as different speakers), we also introduce two corresponding training sets: VoxVietnam-T-GI and VoxVietnam-T-SI. The statistics for each subset are described in Table \ref{tab:subsets-stats}.

\begin{table}[]
\centering
\caption{Statistics of VoxVietnam and Vietnam-Celeb subsets}
\label{tab:subsets-stats}
\begin{tabular}{|l|rr|}
\hline
\textbf{Subset}    & \textbf{\# of Speakers} & \textbf{\# of Utterances} \\ \hline\hline
Vietnam-Celeb-T    & 880                     & 82,907                    \\
VoxVietnam-T       & 1,256                   & 161,457                   \\
VoxVietnam-T-\textit{small} & 879                     & 83,000                    \\
VoxVietnam-T-\textit{noisy} & 1,256                     & 384,347                    \\
VoxVietnam-T-GI    & 2,191                   & 161,457                   \\
VoxVietnam-T-SI    & 2,316                   & 161,457                   \\ \hline
VoxVietnam-E       & 150                     & 26,523                    \\
VoxVietnam-H       & 150                     & 26,523                    \\ \hline
\end{tabular}
\end{table}

The test pairs in the easy subset VoxVietnam-E are chosen randomly. The hard subset VoxVietnam-H is injected with a few pairs of utterances of different genres from the same speaker, and utterances of different speakers with high cosine similarity as negative pairs to highlight the challenges posed by multi-genre scenarios. The labels of all the test pairs in VoxVietnam-E and VoxVietnam-H are manually verified by humans.

\subsection{Experimental Results}
The experimental results are illustrated in Table \ref{tab:eer-results}. Firstly, we evaluate the performance of the ECAPA-TDNN model trained on single-genre (Vietnam-Celeb-T) and multi-genre (VoxVietnam-T) on multi-genre test utterances. It can be seen that single-genre training performs poorly in multi-genre cases, with an absolute increase of up to 0.88\% in EER compared to multi-genre training. The sampled training set VoxVietnam-T-\textit{small} also helps to slightly improve the performance compared to the Vietnam-Celeb-T of comparable size. 

Moreover, we also investigate the efficacy of the proposed dataset when combined with Vietnam-Celeb (Vietnam-Celeb-T + VoxVietnam-T), or when used as a supportive fine-tuning resource (Vietnam-Celeb-T ft. VoxVietnam-T). The proposed VoxVietnam also shows a significant improvement when combined with Vietnam-Celeb for training or fine-tuning, with an absolute decrease of 1.37\% and 0.98\% on the easy evaluation set, 0.88\% and 0.39\% on the hard evaluation set.

The single-genre and multi-session speakers are also of significant importance. They only show a slight relative decrease of 2.97\% on average when compared to training with true multi-genre speakers, while being competitive to training with single-genre speakers in Vietnam-Celeb-T. This is an advantage since the collection of true multi-genre speakers in low-resource languages is a challenge owing to the scarcity of public media sources.
\begin{table}[]
\centering
\caption{EER (\%) results on different train and test setup}
\label{tab:eer-results}
\begin{tabular}{|l|rr|}
\hline
\backslashbox[45mm]{\textbf{Train}}{\textbf{Test}} & \textbf{VoxVietnam-E} & \textbf{VoxVietnam-H} \\ \hline\hline
Vietnam-Celeb-T    & 14.17 & 22.69 \\
VoxVietnam-T   & 13.29  & 21.94  \\
VoxVietnam-T-\textit{small} & 13.40 & 22.22 \\
VoxVietnam-T-\textit{noisy} & 14.91 & 24.14 \\
Vietnam-Celeb-T + VoxVietnam-T & \textbf{12.80}       & \textbf{21.81}       \\
Vietnam-Celeb-T ft. VoxVietnam-T  &   13.19           & 22.30  \\
VoxVietnam-T-GI & 13.32 & 22.39 \\
VoxVietnam-T-SI & 14.05 & 22.94 \\ 

\hline
\end{tabular}
\end{table}

Experiments are also done on the noisy training set to evaluate the performance gap when speaker cleansing is utilised. It can be seen that the cleansing step helps to improve the absolute performance by 1.21\% on average. Finally, we conduct within-genre and cross-genre experiments to study the performance in different genres. In a within-genre setup, the training and test utterances are of the same genre. To avoid bias towards the number of training data, the same number of speakers and utterances per speaker is sampled for every of the three genres. The results are shown in Figure \ref{fig:within-cross-genre}.
\begin{figure}
    \centering
    \includegraphics[width=0.75\linewidth]{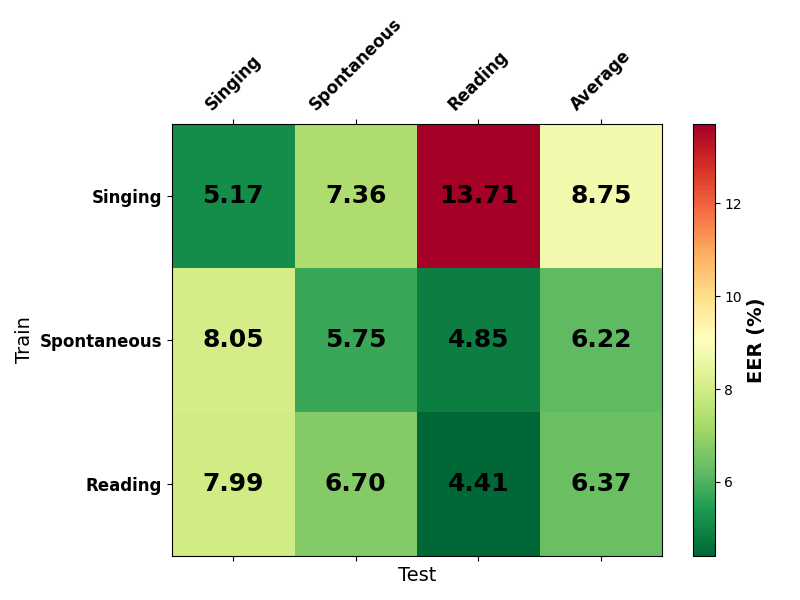}
    \caption{Within-genre and Cross-genre evaluations.}
    \label{fig:within-cross-genre}
\end{figure}

In most cases, the cross-genre is worse than the within-genre evaluations. Moreover, the best performance is obtained when the enrolment is in speech-like genres (reading and spontaneous), whereas the worst performance appears in complicated genres like singing. These results align with the proven observation in \cite{li2022cn}, where the simpler the enrolment utterance condition is, the better the overall verification performance. This is also an advantage for verification models since speakers tend to enrol in stable, ambient environments with careful pronunciation.

\section{Conclusions}
To summarise, we made three major contributions in this paper: (i) proposed a novel and scalable pipeline of collecting, processing, and cleansing to construct a large-scale, language-independent speaker recognition dataset; (ii) introduced and freely published the largest-scale and first multi-genre dataset for Vietnamese speaker recognition; and (iii) conducted extensive experiments to study to effects of multi-genre training. The experiments show that multi-genre test utterances can lead to an absolute degrade of 0.8\% in EER. Moreover, the experiments also express the efficacy of VoxVietnam for fine-tuning or when combined with another dataset, with an absolute improvement of 0.9\% on average. The dataset is publicly available on HuggingFace\footnote{https://huggingface.co/datasets/hustep-lab/VoxVietnam-Dataset} for researchers to conduct further experiments on multi-genre challenges in the Vietnamese language, and to develop more robust recognition models. Future works on VoxVietnam will focus on expanding the diversity of utterance genre, and balance the genre and gender among the speakers.

% \end{thebibliography}
\newpage
\bibliographystyle{IEEEtran}
\bibliography{mybib}
\end{document}